\documentclass[twocolumn,superscriptaddress,floatfix,preprintnumbers,prd ,nofootinbib,hyperref]{revtex4-2}
\usepackage[colorlinks=true,breaklinks=true]{hyperref}
\usepackage{comment}
\usepackage[utf8]{inputenc}
\hypersetup{allcolors=[rgb]{0.0 0.0 0.6},linkcolor=[rgb]{0.75 0.05 0.05}}
\usepackage{mathtools}
\usepackage[dvipsnames]{xcolor}
\usepackage{float}
\usepackage{letltxmacro}
\LetLtxMacro{\oldcite}{\cite}
\renewcommand{\cite}[1]{\mbox{\oldcite{#1}}}

\usepackage{orcidlink}

\long\def\exclude#1{}

\bibpunct{[}{]}{,}{n}{}{,}

\begin{document}

\title{Axion-sourced fireballs from supernovae}

\author{Melissa Diamond \orcidlink{0000-0003-1221-9475}} 
\affiliation{Arthur B. McDonald Canadian Astroparticle Physics Institute,
Queens University, Kingston, Ontario, Canada K7L 3N6}

\author{Damiano F. G. Fiorillo
\orcidlink{0000-0003-4927-9850}}
\affiliation{Niels Bohr International Academy \& DARK, Niels Bohr Institute,
University of Copenhagen, Blegdamsvej 17, 2100 Copenhagen, Denmark}

\author{Gustavo Marques-Tavares
\orcidlink{0000-0002-1861-7936}}
\affiliation{Maryland Center for Fundamental Physics, Department of Physics,
University of Maryland, College Park, MD 20742, U.S.A.}

\author{Edoardo Vitagliano
\orcidlink{0000-0001-7847-1281}}
\affiliation{Racah Institute of Physics, Hebrew University of Jerusalem, Jerusalem 91904, Israel}

\date{\today}

\begin{abstract}

New feebly interacting particles would emerge from a supernova core
with 100-MeV-range energies and produce $\gamma$-rays by subsequent
decays. These would contribute to the diffuse cosmic $\gamma$-ray
background or would have shown up in the Solar Maximum Mission (SMM)
satellite from SN~1987A. However, we show for the example of
axion-like particles (ALPs) that, even at distances beyond the
progenitor star, the decay photons may not escape, and can instead form a
fireball, a plasma shell with $T\alt1$~MeV. Thus, existing arguments do
not exclude ALPs with few 10 MeV masses and a two-photon coupling of a
few $10^{-10}~{\rm GeV}^{-1}$. However, the energy would have showed up in
sub-MeV photons, which were not seen from SN~1987A in the Pioneer Venus
Orbiter (PVO), closing again this new window. A careful re-assessment
is required for other particles that were constrained in similar ways.
\end{abstract}

\maketitle

\section{Introduction}

The core of a collapsing star is one of the hottest and densest regions in the Universe. Therefore, it can be a factory of high-energy (around 100 MeV) particles beyond the standard model, in particular, feebly interacting particles (FIPs), such as sterile neutrinos, dark photons, new scalars, QCD axions and axion-like particles, and many others. It comes as no surprise, nowadays, that---despite the small number of neutrino events detected at several neutrino experiments---supernova~1987A (SN~1987A) represented a bonanza for bounds on FIPs. A renowned example is the SN~1987A energy-loss bound~\cite{Mayle:1987as,Mayle:1989yx,Turner:1987by}, that limits the luminosity of a novel particle $\phi$ to be smaller than the neutrino luminosity, $L_\phi\lesssim L_\nu$, evaluated at 1 second after the bounce~\cite{Raffelt:1996wa,Raffelt:2006cw}.
While a large body of literature is dedicated to light particles, the last decade has seen an ever-growing interest in the high-mass part of FIP parameter space, above $1\,\rm keV$, that could have an impact on cosmology~\cite{Cadamuro:2011fd,Depta:2020zbh,Kelly:2020aks,Langhoff:2022bij}, astrophysical transients~\cite{Diamond:2021ekg,Caputo:2021kcv}, and play the role of dark matter mediator~\cite{Pospelov:2007mp,Knapen:2017xzo}.

Heavy FIPs can be probed down to luminosities much smaller than $L_\nu$. Depending on their lifetime and decay channels, they can travel for distances that are either smaller or larger than the progenitor radius. If the mean-free path against decay is small enough, FIPs can decay in the mantle of the progenitor, lighting up the SN~\cite{Falk:1978kf}. Since the explosion energy of SNe is much smaller than the energy released in neutrinos, the radiative decay of FIPs to charged leptons and photons must be suppressed~\cite{Sung:2019xie,Caputo:2021rux}. This argument, applied to low energy SNe, limits the lifetime of radiatively decaying particles even further~\cite{Caputo:2022mah}. If the mean-free path against the decay is larger (or if the FIP decays to neutrinos), the decay product of FIPs produced by SN~1987A could have been seen \textit{directly} in the detectors on-line during the explosion~\cite{Chupp:1989kx,Kolb:1988pe,Oberauer:1993yr,Jaeckel:2017tud}. For example, the decay to 100-MeV neutrinos would have been seen in Cherenkov detectors, so the lack of such events gives constraints superseding the energy-loss bound by one order of magnitude in coupling~\cite{Fiorillo:2022cdq}.

\begin{figure}[t!]
    \centering
    \includegraphics[width=0.55\textwidth]{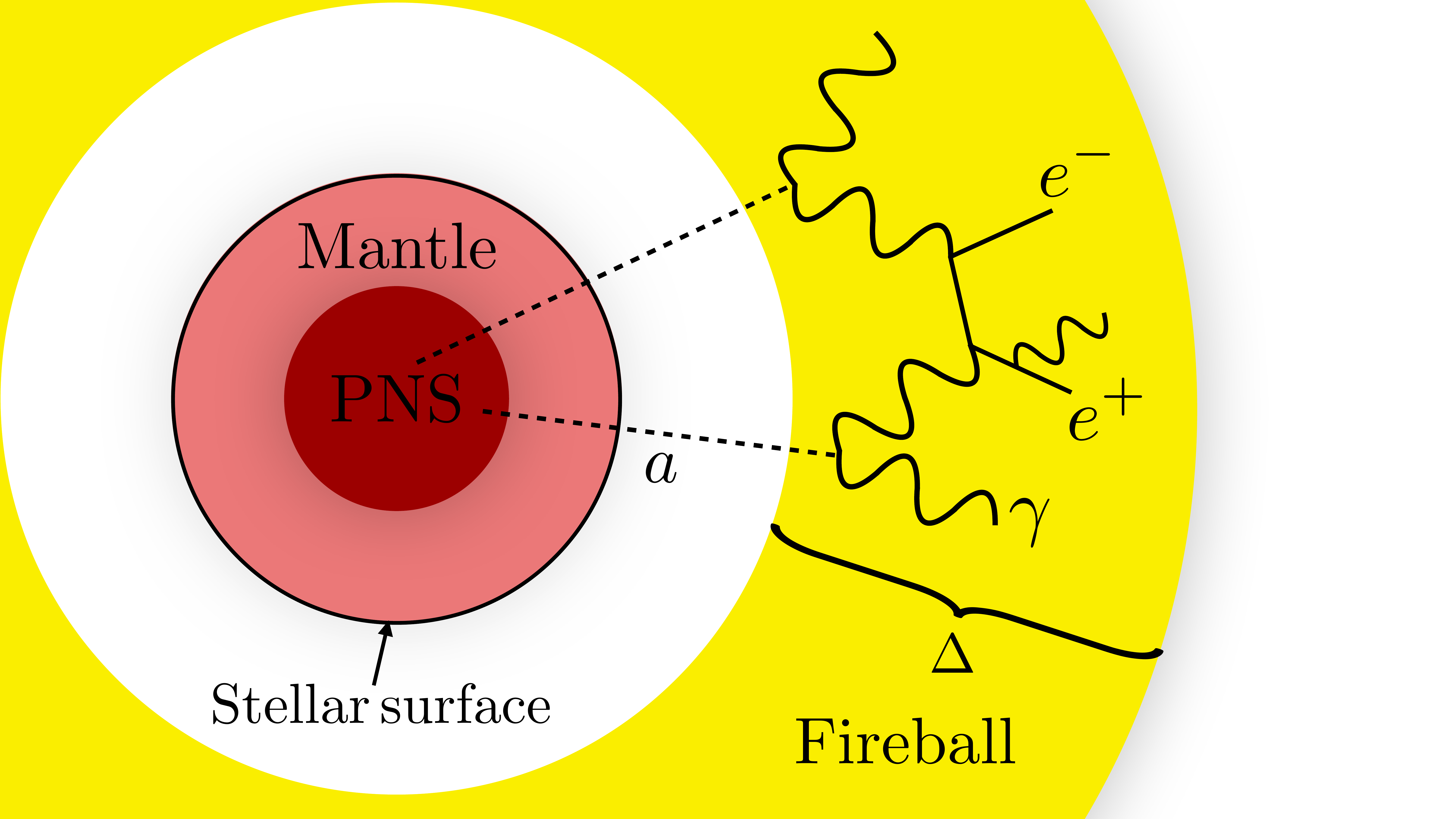}
    \caption{(Out-of-scale) Schematic representation of the fireball sourced by axions from a supernova. Axions are produced in the proto-neutron star, travel outside of the progenitor star, and decay exterior to its surface. The produced $\gamma$-rays, rather than conserving their spectrum, form a fireball.}
    \label{fig:disegno}
\end{figure}

\begin{figure*}[t!]
    \centering
    \includegraphics[width=\textwidth]{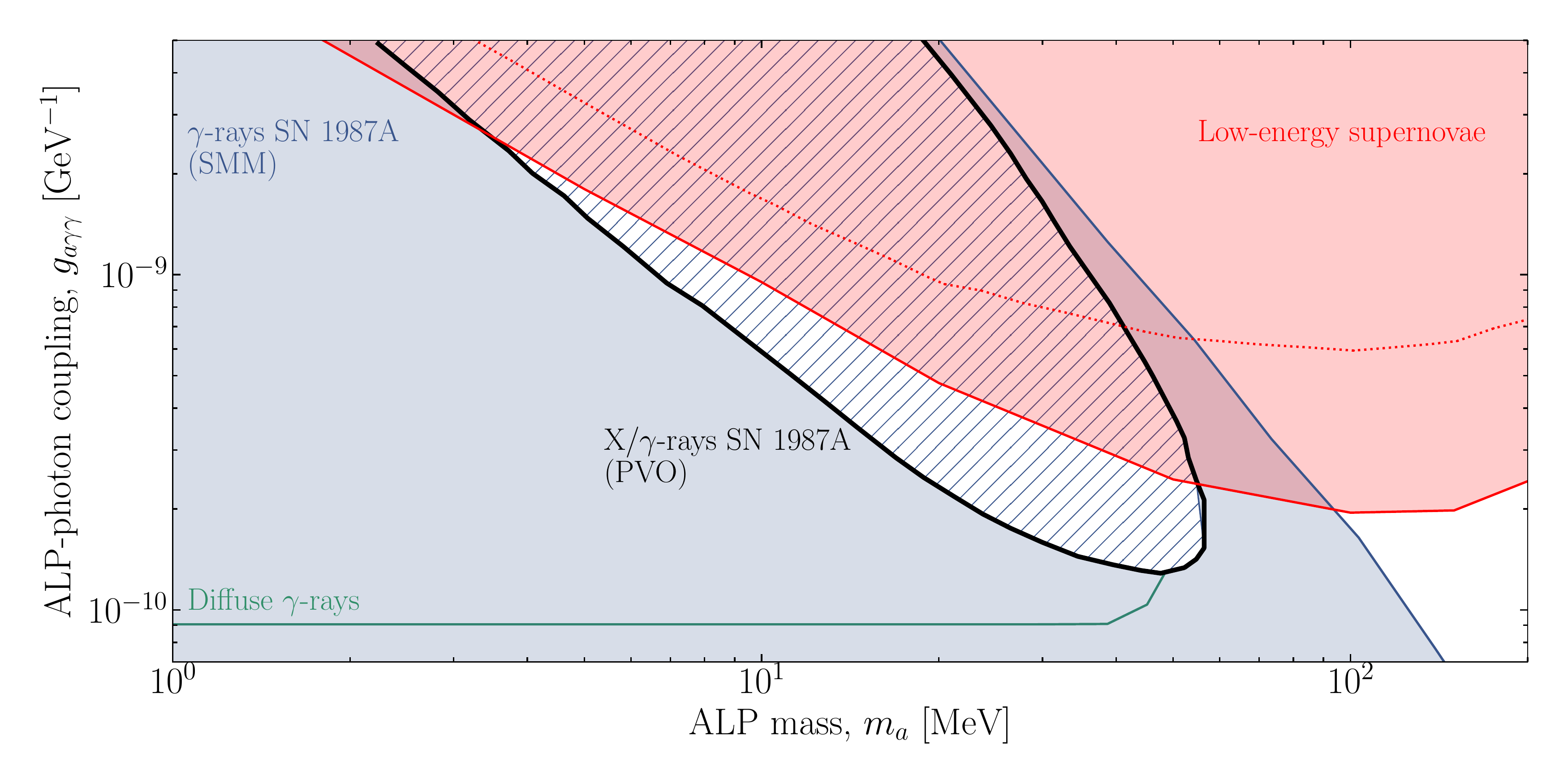}
    \caption{Region of fireball formation in the axion mass and coupling space (solid black line). We also show  bounds from low-energy supernovae (red)~\cite{Caputo:2022mah}, $\gamma$-rays due to axion decays from SN~1987A (blue)~\cite{Jaeckel:2017tud,Hoof:2022xbe}, and diffuse supernova background of $\gamma$-rays from axion decays (green)~\cite{Caputo:2021rux,Caputo:2022mah}. The constraints from low-energy supernovae are the only ones unaffected by the fireball formation.
    Bounds from $\gamma$-ray decay at the Solar Maximum Mission and from the diffuse supernova background should be reevaluated in the fireball formation region. On the other hand, we find that the entire region, which we hatch in blue, is robustly excluded by the Pioneer Venus Orbiter observations.}
    \label{fig:final_bounds}
\end{figure*}
The focus of this paper will be on heavy axion-like particles (hereafter axions) that can decay to photons through the coupling $-\frac{1}{4}g_{a\gamma\gamma}a F\Tilde{F}$. 
The daughter photons would have contributed to the diffuse cosmic $\gamma$-ray background~\cite{Fermi-LAT:2014ryh,Calore:2020tjw,Caputo:2021rux} and, for SN~1987A, would have showed up in the Gamma-Ray Spectrometer on the Solar Maximum Mission satellite~\cite{forrest1980gamma}. The latter measured the fluence in the interval $4.1-100\,\, \rm MeV$ during a $223.2\,\rm s$ interval coincident with the SN~1987A neutrino burst. Its observations have been used to constrain radiatively decaying particles by looking for $\gamma$-ray emission~\cite{Jaeckel:2017tud,Caputo:2021rux,Ferreira:2022xlw,Hoof:2022xbe}. 

For such constraint to apply, the photons produced in the axion decay should have been in the energy interval detected by SMM~\cite{Chupp:1989kx}. If axions with a mass $m_a\gtrsim 1 \,\, \rm MeV$ were produced in SN~1987A, after escaping from the progenitor they would have decayed to $\gamma$-ray photons with energy around $100\,\, \rm MeV$. If the density of such photons was sufficiently large, they would rapidly produce a fireball, creating a plasma of electrons, positrons and photons, as sketched in Fig.~\ref{fig:disegno}. Several processes rapidly drive the plasma to a much lower temperature. After this rapid thermalization, the plasma would follow the evolution of a ``standard'' fireball~\cite{Piran:1999kx,Meszaros:2006rc}, though with a much smaller initial temperature and, potentially, negligible baryon load.\footnote{The decay of heavy axions was proposed as a mechanism to produce the fireball sourcing gamma-ray bursts or as a SN catalyzer in Ref.~\cite{Berezhiani:1999qh}, but the axion luminosity needed to source such kind of fireball requires couplings that are excluded.} The gas would first expand adiabatically, converting the temperature into bulk momentum, and then expand freely. The photons produced in axion decays from SN~1987A would have had an energy $E_\gamma\lesssim 1\,\, \rm MeV$. Most of these photons would not have been energetic enough to be detected by SMM, though we are still able to put constraints using the data of the Pioneer Venus Orbiter Satellite (PVO)~\cite{PVO}, which had an energy window $0.2-2\,\rm MeV$. Our main results are collected in Fig.~\ref{fig:final_bounds}. Our new PVO bound covers the region of the parameter space carved by the fireball formation and previously thought to be excluded by SMM.

The paper is structured as follows. In Section~\ref{sec:formation} we describe the first stages of the fireball, from its formation to thermalization. In Section~\ref{sec:expansion} we analyze the period of expansion. Section~\ref{sec:bounds} and Section~\ref{sec:diffuse} are dedicated to the impact of the fireball formation on  respectively the SN 1987A and diffuse $\gamma$-ray bounds, as well as to the new bounds we place on the axion parameter space. Finally, we devote Section~\ref{sec:discussion} to our conclusions.

\section{Formation of the fireball}\label{sec:formation}

Axions are produced in the protoneutron star (PNS) at the center of the supernova, mainly via Primakoff emission and photon-photon coalescence. Their spectrum is parameterized in Refs.~\cite{Jaeckel:2017tud, Hoof:2022xbe}. Both papers account for Primakoff, and neglect coalescence, which is reasonable at light masses. Since we want to compare the region of fireball formation with the bounds drawn in Refs.~\cite{Jaeckel:2017tud, Hoof:2022xbe}, we self-consistently use their expressions; coalescence would increase the total number of photons injected and enhance the possibility of fireball formation. Further, we are mostly interested in masses below~60~MeV, as confirmed by our results, where coalescence should have little impact (see e.g. the supplemental material of Ref.~\cite{Caputo:2022mah}).

The axion spectrum is extracted from Ref.~\cite{Hoof:2022xbe} as
\begin{equation}\label{eq:hoof}
\frac{dN_a}{dE_a}=C_2\frac{E_a^2}{\exp(E_a/T_\mathrm{eff})-1}\sigma_0(E_a,g_{a\gamma}, \kappa_s,m_a).
\end{equation}
The numerical values for the parameters $C_2$, $\kappa_s$, $T_\mathrm{eff}$ are all taken from Ref.~\cite{Hoof:2022xbe}; the expression for the Primakoff cross section for massive particles, $\sigma_0(E_a,g_{a\gamma},\kappa_s,m_a)$, is taken from Ref.~\cite{Jaeckel:2017tud}.

Only those axions which decay outside of the progenitor, with a radius of approximately $R=3\times10^{12}$~cm, contribute to the photons which are detected at Earth. Therefore, the total energy injected by the axions is
\begin{equation}\label{eq:tot_energy}
    \mathcal{E}=\int E_a \frac{dN_a}{dE_a} e^{-R/\ell(E_a)} dE_a,
\end{equation}
where the decay length is 
\begin{equation}
    \ell(E_a)=\frac{\sqrt{E_a^2-m_a^2} 64\pi}{g_{a\gamma}^2 m_a^4}.
\end{equation}
We can similarly determine the total number of axions injected $\mathcal{N}$ and the total radial momentum injected $\mathcal{P}$. Notice that SN 1987A had an atypical small radius. For a larger progenitor radius, as could be the case of a future supernova---for example, Betelgeuse has a radius about one order of magnitude larger than SN 1987A progenitor, Sanduleak -69 202~\cite{2020ApJ...902...63J}---a smaller fraction of axions decay outside of the mantle.

\subsection{Geometry of the fireball}

The massive axions propagate with a speed close to the speed of light, approximately radially, and decay substantially far from the center of the supernova. The decay is in reality a continuous process; however, most of the energy from the axion decay is injected at a distance of the order of the decay length. The latter is larger than the typical thickness of the axion shell propagating from the supernova at small couplings and masses. Based on these facts, we can schematically model the propagation as a shell of axions, with a thickness of the order of $\Delta_1\simeq3\,\rm s$, the typical timescale over which they are emitted from the supernova.

The typical decay radius for axions that decay outside the progenitor is determined by averaging over both the energy distribution and the radius of decay distribution, $e^{-(r_\mathrm{dec}-R)/\ell(E_a)} dr_\mathrm{dec}/\ell(E_a)$, where $r_\mathrm{dec}>R$; this leads to the average radius
\begin{equation}
    r=R+\frac{\int dE_a \ell(E_a) e^{-R/\ell(E_a)}\frac{dN_a}{dE_a}}{\int dE_a  e^{-R/\ell(E_a)}\frac{dN_a}{dE_a}}.
\end{equation}
For simplicity, here and henceforth we will indicate the average of a quantity $x$ over the energy fluence of the axions decaying outside the progenitor as
\begin{equation}
    \langle x \rangle=\frac{\int dE_a x e^{-R/\ell(E_a)}\frac{dN_a}{dE_a}}{\int dE_a  e^{-R/\ell(E_a)}\frac{dN_a}{dE_a}}.
\end{equation}

To estimate the thickness of the photon shell, we determine the quadratic dispersion in the radius of the photons evaluated at the time $\bar{t}=\langle \tau(E_a)+R/v(E_a)\rangle=\langle(\ell(E_a)+R)/v(E_a)\rangle$, where $v(E_a)=\sqrt{1-m_a^2/E_a^2}$; this is the mean time at which the bulk of the axions outside the progenitor decay. If the axion is produced in the PNS at a time $t_i$, and decays at time $t_\mathrm{dec}$, the photon position at time $t$ is
\begin{equation}
    r_\gamma=v(E_a)\left[t_\mathrm{dec}-t_i\right]+t-t_\mathrm{dec}.
\end{equation}
The assumption that photons propagate radially is valid for axions with masses much below $\bar{E}_a\simeq 100\, \rm MeV$, the typical axion energy. We will be interested in typical masses below $60\, \rm MeV$, since at larger masses we find no fireball formation. Therefore, accounting for the angle with which the photons are emitted in the decay, typically of the order $m_a/\bar{E}_a$, would account for $10\%$ corrections to the average radius of photon emission. Without loss of generality, we shift the origin of time and define $t_i$ to be uniformly distributed between $-\Delta_1/2$ and $\Delta_1/2$, where the time window is given a representative value $\Delta_1=3\,\rm s$; with this choice the average value of $t_i$ vanishes. At a fixed energy, the decay time is averaged over the exponential distribution $e^{-\left[t_\mathrm{dec}-R/v(E_a)\right]/\tau(E_a)} dt_\mathrm{dec}/\tau(E_a)$ for $t_\mathrm{dec}>R/v(E_a)$. The average position of a photon at time $t$ therefore is
\begin{equation}
    \langle r_\gamma \rangle=t+\langle (v(E_a)-1)\left(\tau(E_a)+\frac{R}{v(E_a)}\right)\rangle.
\end{equation}
Evaluating this at $\bar{t}$, we find, as we should, our previous estimate for the fireball radius,
\begin{equation}
    \langle r_\gamma \rangle=\langle v(E_a)\left(\tau(E_a)+\frac{R}{v(E_a)}\right)\rangle.
\end{equation}
We now compute the average of the squared radius and define the thickness as
\begin{equation}
    \Delta^2=\langle r_\gamma^2\rangle-\langle r_\gamma\rangle^2.
\end{equation}
Expanding the averages over the exponential distribution, this is
\begin{align}
        \Delta^2=\,&\frac{\Delta_1^2 \langle v(E_a)^2\rangle}{12} \nonumber\\ \nonumber &+\langle(1-v)^2\left(\frac{R^2}{v(E_a)^2}+\frac{2R\tau(E_a)}{v(E_a)}+2\tau(E_a)^2\right)\rangle\\ &-\left[\langle\tau(E_a)+\frac{R}{v(E_a)}\rangle-\langle v(E_a) \tau(E_a)+R\rangle\right]^2.
\end{align}
The first term corresponds to the average thickness of the axion shell, while the other terms correspond to the thickness induced by the delay between the slower axions yet to decay and the photons. Notice that for ultra-relativistic axions the last two terms become arbitrarily small, since axions and photons move with arbitrarily close velocities. Therefore, its order of magnitude for strongly boosted axions, as we can approximately consider for masses below $60\, \rm MeV$ that are of interest here, is $r m_a^2/\bar{E}_a^2$. The thickness of the photon shell will therefore be determined by the largest among the thickness of the axion shell and the approximate spread $r m_a^2/\bar{E}_a^2$. Notice that in the thickness of the shell the corrections due to the non-collinear emission of photons in axion decay might lead to corrections by factors of order $1$, since these are suppressed by one factor of $m_a^2/\bar{E}_a^2$ and must be compared with terms of the same order of magnitude. On the other hand, we will find later that the thickness of the shell in the end never enters the final results presented here. Therefore, we stick to this simple estimate which might be corrected by factors of order unity, since none of our results are affected by it.

\subsection{Thermalization via pair production}

If the photon density is large enough, it can form a plasma by producing pairs of $e^\pm$. The pair production process (and subsequent Compton and Rutherford scattering) leads to kinetic equilibrium, with both $\gamma$ and $e^\pm$ acquiring thermal spectra. Since pair production is not able to change the total number of particles, chemical equilibrium cannot be fully reached, and both species develop a common temperature $T_i$ and (negative) chemical potential $-\mu$. Furthermore, they behave as a tightly coupled fluid with a bulk Lorentz factor $\gamma$. 

In the following, we will refer with primes to lab-frame quantities, and without primes to plasma-frame quantities. If the number density of photons $n'_0=2\mathcal{N}/4\pi r^2 \Delta$ is too low, they are not able to produce pairs of electrons and positrons. 
The condition for pair-production reactions to be efficient is
\begin{equation}\label{eq:pairprodequi}
    n'_0 \sigma_{\gamma\gamma\to e^+e^-}\Delta\gg1,
\end{equation}
where $\sigma_{\gamma\gamma\to e^+e^-}$ is the pair production cross section, to be evaluated at the typical photon energies in the plasma frame, where the photons are isotropic. The cross section is therefore~\cite{Akhiezer:1986yqm}
\begin{align}
&\sigma_{\gamma\gamma\to e^+e^-}=\frac{\pi \alpha^2}{E_\gamma^2}\left[\left(2+\frac{2 m_e^2}{E_\gamma^2}-\frac{m_e^4}{E_\gamma^4}\right) \right.
\nonumber
\\
&\times\left.\log\left|\frac{E_\gamma}{m_e}+\sqrt{\frac{E_\gamma^2}{m_e^2}-1}\right|-\sqrt{1-\frac{m_e^2}{E_\gamma^2}}\left(1+\frac{m_e^2}{E_\gamma^2}\right)\right],
\end{align}
where $E_\gamma$ is the center-of-mass energy of any of the two colliding photons. We average this cross section over a uniform distribution of the pitch angle between the two incoming photons, with all the photons having a constant energy $m_a/2$ in the comoving frame. In reality, there will be photons from axions moving with different speeds with a relative angular distribution different from the isotropic one and with a relative boost with respect to one another; however, these small corrections typically change negligibly the shape or normalization of our curves corresponding to fireball formation, due to the large power of the coupling appearing in these curves (see below).

If Eq.~\eqref{eq:pairprodequi} is satisfied, a population of electrons and positrons is nearly immediately established in the plasma. Because the typical energies at this stage are much larger than the electron mass, the energy and number of particles are equally shared among the two populations of electrons and positrons and the population of photons. The corresponding fluid has  equation of state $p=\rho/3$, where $p$ is the fluid pressure and $\rho$ is the fluid energy density. Therefore, we may deduce its bulk velocity taking the ratio of the conservation of energy density (see, e.g., Ref.~\cite{Weinberg:1972kfs}, Chap.~2)
\begin{equation}\label{eq:energy_conservation}
    \frac{4\gamma^2-1}{3}\rho=\frac{\mathcal{E}}{4\pi r^2\Delta}
\end{equation}
and momentum density
\begin{equation}\label{eq:momentum_conservation}
    \frac{4}{3}\gamma^2 \rho v=\frac{\mathcal{P}}{4\pi r^2 \Delta}.
\end{equation}
Here, $v$ is the bulk velocity and $\gamma$ is the bulk Lorentz factor. After taking the ratio, we obtain
\begin{equation}\label{eq:bulk_velocity}
    v=\frac{2\mathcal{E}}{\mathcal{P}}-\sqrt{\frac{4\mathcal{E}^2}{\mathcal{P}^2}-3}.
\end{equation}
We can also deduce the initial temperature of this plasma using the conservation of the total number of particles, which at this stage is valid and yields
\begin{equation}\label{eq:number_conservation}
    \gamma n=\frac{2\mathcal{N}}{4\pi r^2\Delta},
\end{equation}
where $n$ is the number density of the fluid. Since at this stage chemical equilibrium is not maintained, the chemical potential of photons, electrons, and positrons is equal to a common (negative) value $\mu_\gamma=\mu_{e^+}=\mu_{e^-}=-|\mu|$, such that $\mu\gg T$. Therefore, we can use the equation of state of a gas of relativistic Boltzmann particles: taking the ratio of Eqs.~\eqref{eq:momentum_conservation} and~\eqref{eq:number_conservation} we obtain the initial temperature of the plasma as
\begin{equation}\label{eq:initial_temperature_plasma}
    T_i=\frac{\mathcal{P}}{8\gamma v \mathcal{N}}.
\end{equation}

\subsection{Thermalization via bremsstrahlung}

\begin{figure}
    \centering
    \includegraphics[width=0.5\textwidth]{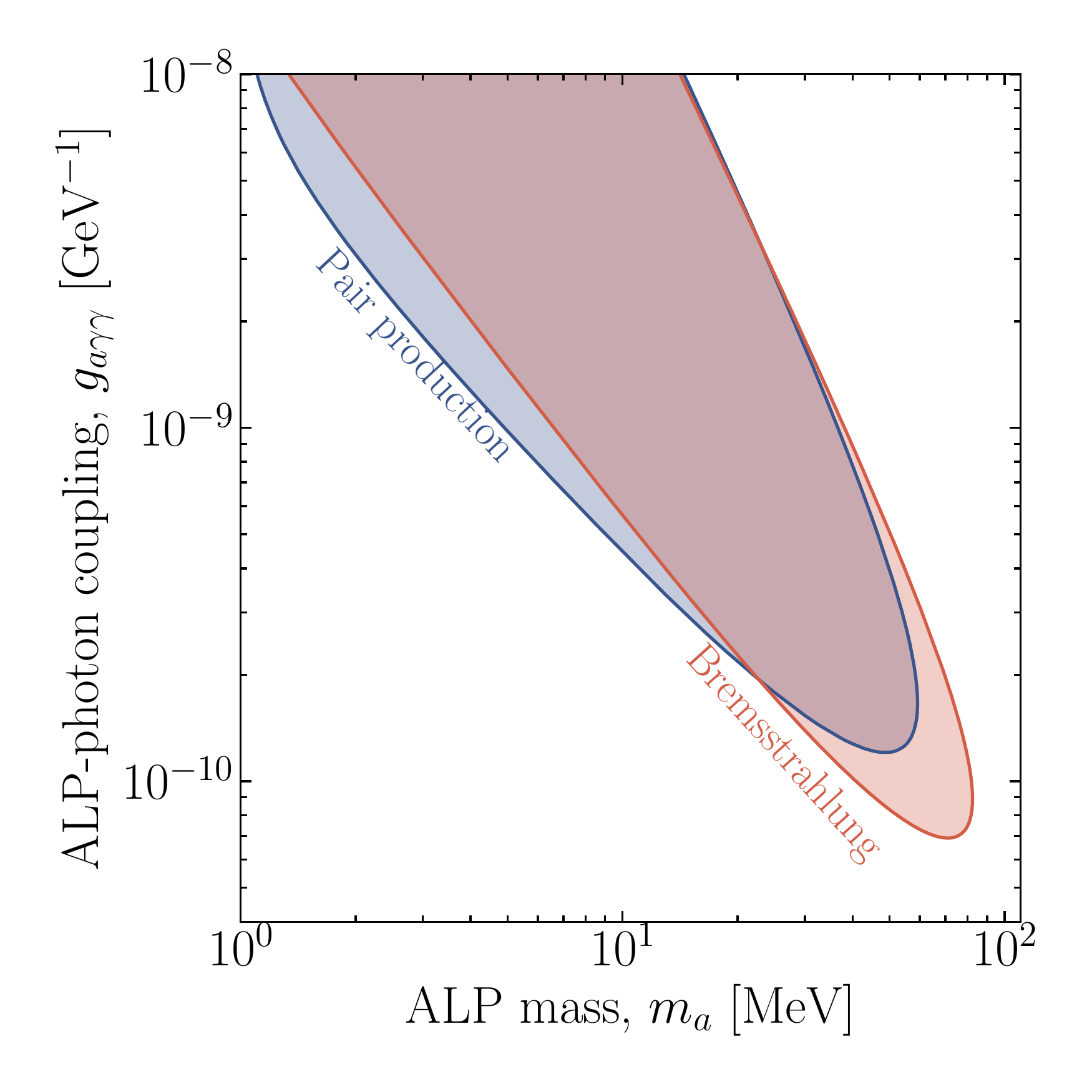}
    \caption{Regions of the parameter space in which the relevant processes for chemical equilibrium are activated. In the blue region, the photon plasma from axion decay is dense enough that pair production equilibrates, leading to the formation of an electron-positron component in the plasma. In the red region, if electrons and positrons are produced, the number density is diluted by bremsstrahlung emission. Thus, in the intersection of the blue and red region, number density dilution via bremsstrahlung is effective.}
    \label{fig:plot_regions}
\end{figure}

With a sufficiently large concentration of $e^\pm$, bremsstrahlung reactions can become fast enough to drive the system towards chemical equilibrium, increasing the particle number. The formation of the fireball requires pair production to equilibrate, while the particle number dilution also entails bremsstrahlung. For both reactions to be fast, axions should be produced in large quantities, with a large fraction decaying outside the progenitor.

The condition for the initial population of electrons and positrons to be equilibrated by bremsstrahlung is
\begin{equation}
\frac{2}{3}\frac{2\mathcal{N}}{4\pi r^2 \Delta} \sigma_{ee\to ee\gamma}(T_i)\Delta \gg1,
\end{equation}
where the factor $2/3$ accounts for the fact that only the electron-positron part of the plasma interacts via bremsstrahlung and $\sigma_{ee\to ee\gamma}(T_i)$ is the cross section for bremsstrahlung from ultrarelativistic electrons and positrons. The total bremsstrahlung cross section diverges, due to the emission of soft photons with negligible energies which, however, do not contribute to the thermalization of the bulk of the plasma. We use as an estimate of the bremsstrahlung cross section the averaged energy emission rate per electron from Eq.~(16) of Ref.~\cite{Diamond:2021ekg}, defined as
\begin{equation}
  \sigma_{ee\to ee\gamma}=\frac{1}{\rho_e n_e}\frac{dE_\gamma}{dt},  
\end{equation}
where $n_e$ is the electron density, $\rho_e$ is the electron energy density, and $dE_\gamma/dt$ is the total energy emitted by bremsstrahlung per unit volume per unit time. For bremsstrahlung, different prescriptions are sometimes used in the literature, such as using the total cross section for emission of a photon with an energy larger than a typical electron energy; but this again leads to very similar results to the criterion used here.

Fig.~\ref{fig:plot_regions} shows, for each axion mass, the minimum value of the coupling for which the energy density of photons is large enough to activate pair production and bremsstrahlung (the latter assuming, of course, that electrons and positrons are produced). For low axion masses, and correspondingly low average energies for the photons, pair production equilibrates at lower coupling than bremsstrahlung, as expected given that bremsstrahlung is suppressed by a higher power of the coupling. At large masses, as soon as electrons and positrons are produced, bremsstrahlung reactions become faster than pair production. The reason is that the average energy per particle, $m_a/2$ in the rest frame, is so large that the pair annihilation cross section is suppressed as $\sigma_{e^+e^-\to\gamma\gamma}\sim \alpha^2/(m_a/2)^2$, whereas the bremsstrahlung cross section is $\sigma_{ee\to ee\gamma}\sim \alpha^3/m_e^2$, up to logarithmic corrections. Therefore, for $m_a\sim 2m_e/\sqrt{\alpha}\sim 10\,\, \rm MeV$, bremsstrahlung is faster than pair production. The subsequent thermalization depends on whether pair production or bremsstrahlung is the faster process. The final state of the plasma is, however, independent of this. Nevertheless, we keep the discussion separate for these two cases for clarity. At large enough couplings and masses, the fireball does not form anymore; the reason is that the fraction of photons decaying outside the progenitor star becomes so low that pair production is not efficient. For this reason, we find no fireball formation for axions heavier than about $60\,\rm MeV$.

\subsection{Final state of the fireball}

\begin{figure*}[t!]
    \centering
    \includegraphics[width=0.3\textwidth]{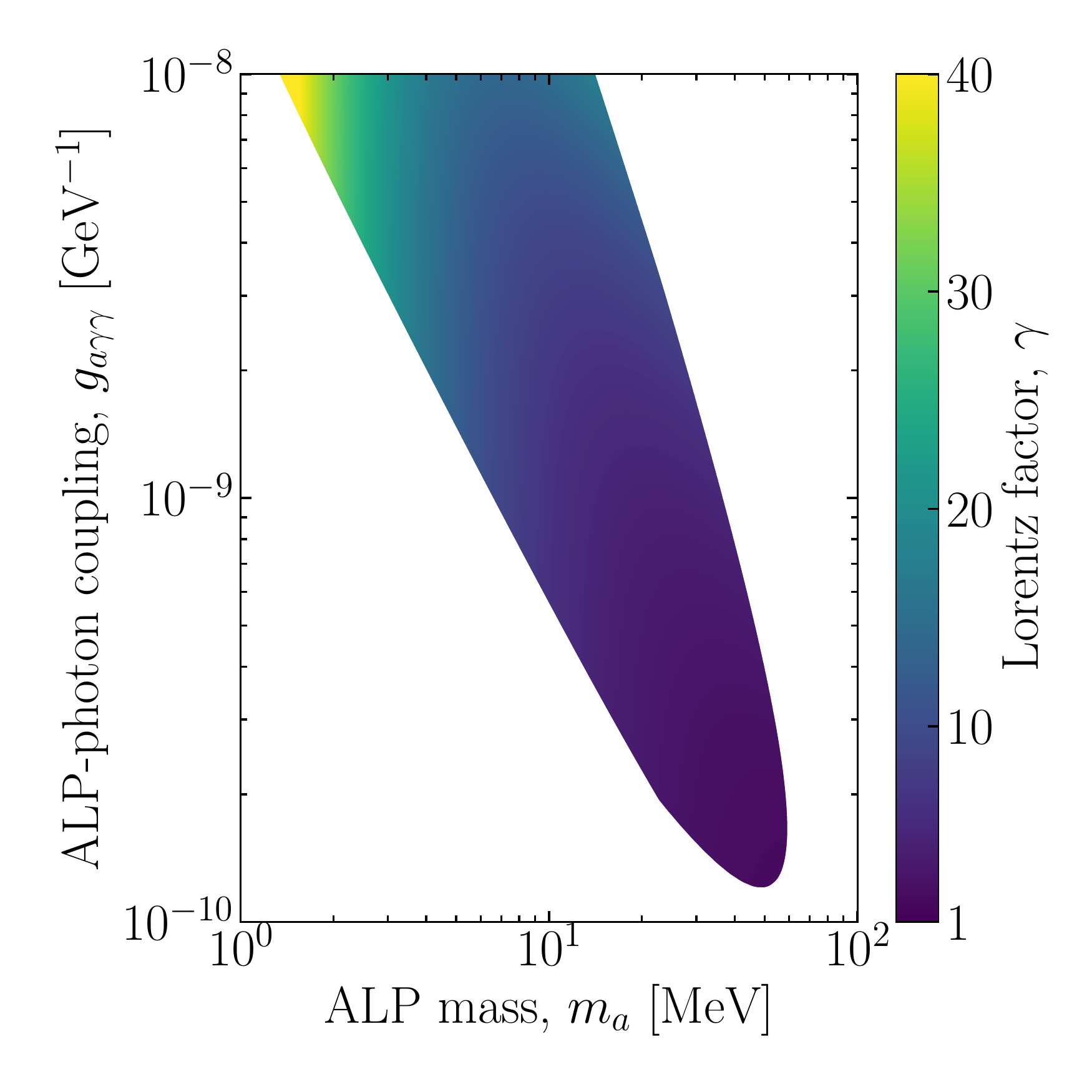}
    \includegraphics[width=0.3\textwidth]{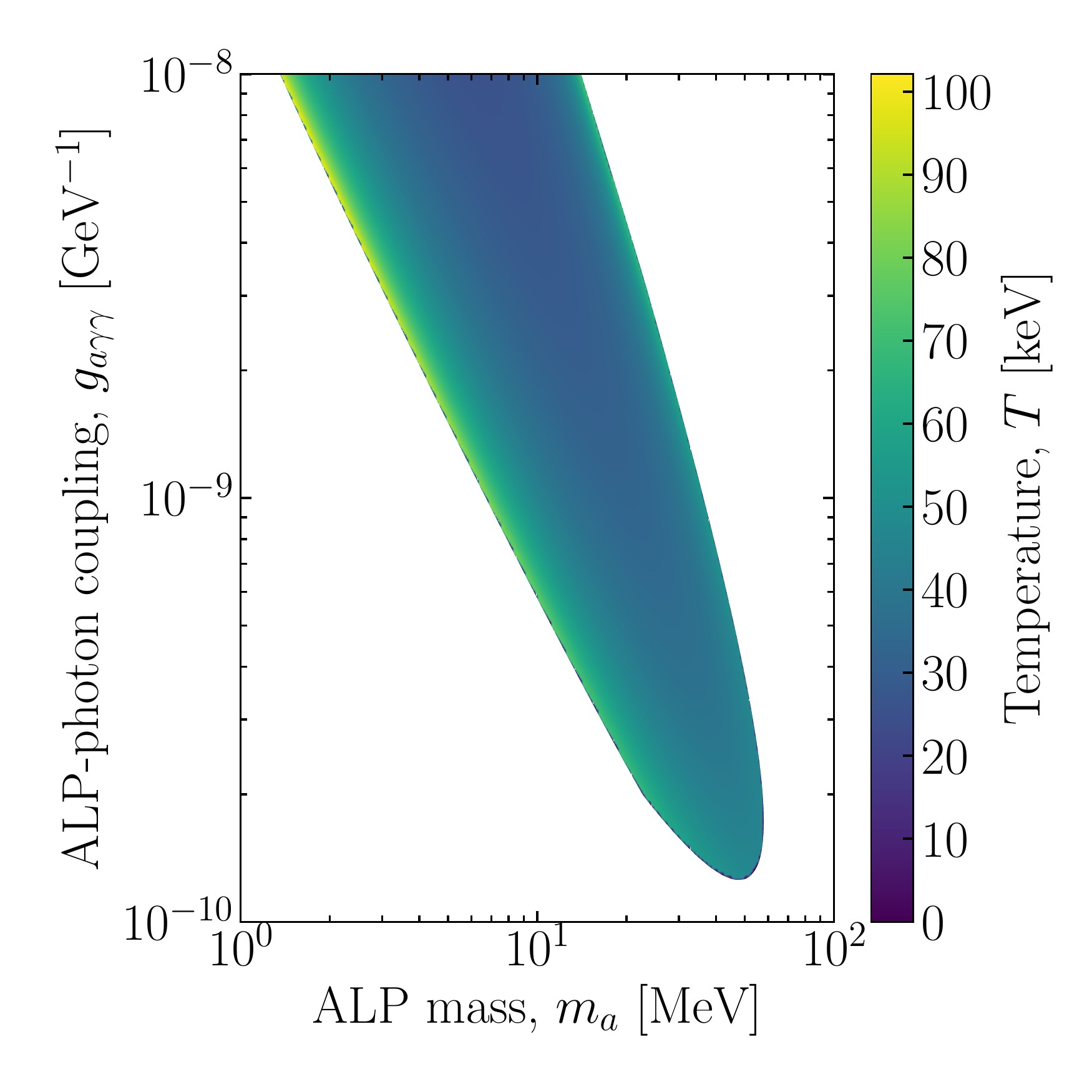}
    \includegraphics[width=0.3\textwidth]{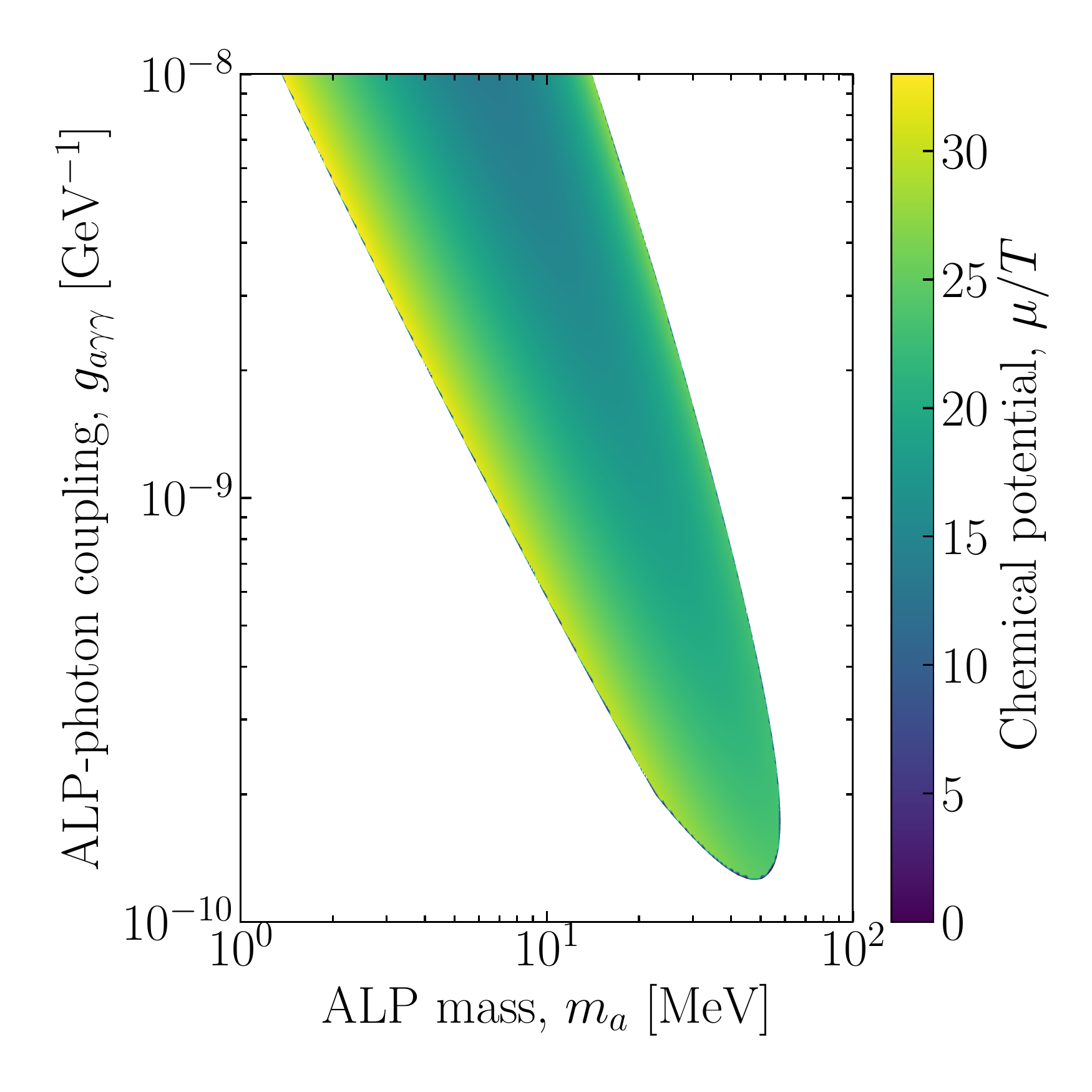}
    \caption{Final state of the plasma. From left to right panel, we show bulk Lorentz factor, rest-frame temperature, and rest frame chemical potential of the plasma, as a function of the axion mass and coupling. All quantities are shown only in the region in which the plasma forms.}
    \label{fig:final_state}
\end{figure*}

Since the total energy released outside of the progenitor, for a benchmark axion mass $m_a=40\,\, \rm MeV$ and coupling $g_{a\gamma\gamma}=2\times10^{-10}$~GeV$^{-1}$, is $\mathcal{E}= 1.22\times10^{50}\,\rm erg$, and the estimated volume of the shell is of order $4\pi r^2 \Delta=3.60\times 10^{38}$~cm$^3$, thermalization tries to drive the plasma towards significantly low temperatures, of the order of $T\simeq \left(15\mathcal{E}/4\pi^3 r^2 \Delta\right)^{1/4}\simeq 192$~eV.\footnote{Since a part of the energy is actually converted in bulk motion because of momentum conservation, this estimate is actually somewhat lowered to $114$~eV.} However, this state is never really reached. Thermalization via bremsstrahlung is interrupted much sooner, since at temperatures smaller than $m_e=0.51\,\, \rm MeV$ the electron population is depleted by the faster pair annihilation.

For $m_a<10\,\, \rm MeV$, in the intersection of the red and blue regions of Fig.~\ref{fig:plot_regions}, pair production is immediately the faster process. The role of pair production is therefore that of maintaining the chemical potential of electrons and photons equal to a common value $\mu_\gamma=\mu_{e^+}=\mu_{e^-}=-|\mu|$. The slower bremsstrahlung reaction (still faster than the timescale associated with the expansion of course) gradually drives this chemical potential towards $0$, with the electrons and positrons emitting photons to dilute the excess average energy per particle. As the chemical potential drops, so does the temperature, since the number of particles is increasing and therefore the average energy per particle decreases. This emission of particles is interrupted when the electron number density is so low that bremsstrahlung cannot proceed any further. Therefore, the final state of the thermalization is a plasma of electrons and positrons with an equal chemical potential $\mu$, with a temperature $T$ and bulk motion $\gamma$ determined by the initial energy and momentum released by the axions, and by the condition that bremsstrahlung is out of equilibrium
\begin{equation}\label{eq:bremsstrahlung_equilibrium}
    \gamma n_e(T,\mu) v_\mathrm{th} \sigma_{ee\to ee\gamma} \Delta=1.
\end{equation}
We use here the non-relativistic expression for the bremsstrahlung cross section, since the number density of electrons and positrons starts to be kinematically suppressed only at temperatures lower than the electron mass. For the same reason, we also account for a factor $v_\mathrm{th}$ corresponding to the typical thermal velocities of the electrons in the plasma frame, of the order of $\sqrt{m_e/T}$. For the full rate of bremsstrahlung interaction, we use the expression~\cite{Diamond:2021ekg}
\begin{equation}\label{eq:nr_brem_cross}
    \sigma_{ee\to ee\gamma}
 v_\mathrm{th}=\frac{64 \alpha^3}{3\sqrt{\pi T m_e^3}}.
 \end{equation}

For $m_a>10\,\, \rm MeV$, in the intersection of the red and blue regions of Fig.~\ref{fig:plot_regions}, after the population of relativistic electrons and positrons is established, bremsstrahlung becomes immediately the faster process. Therefore, in this intermediate stage, the chemical potentials of electrons and photons are not bound to be equal. On the contrary, electrons and positrons start to radiate photons by bremsstrahlung rapidly in order to drive the chemical potential of the photons to zero. In doing this, they dilute the average energy and reduce the temperature of the plasma. Therefore, the initially equal chemical potential of electrons and photons now are unbalanced, with $|\mu_\gamma|<|\mu_{e^+}|=|\mu_{e^-}|$.

This state of affairs is interrupted once the temperature of the plasma drops below $5\,\, \rm MeV$, at which point pair production becomes again the dominant reaction as we have seen above. Therefore, there is a second stage at which pair production rapidly drives the chemical potentials of electrons and photons, which had been unbalanced by bremsstrahlung, to a common value intermediate between the two, so now $\mu_\gamma=\mu_{e^+}=\mu_{e^-}=-|\mu|$. At this point, the subsequent evolution is identical to the case $m_a<10\,\, \rm MeV$: pair production is so rapid that it maintains the chemical potentials equal at all time, and bremsstrahlung slowly dilutes the number density by trying to bring $\mu$ to 0, and the chemical equilibration is interrupted once the bremsstrahlung freezes out, namely when the condition in Eq.~\eqref{eq:bremsstrahlung_equilibrium} is reached.

Thus, the cases $m_a<10\,\, \rm MeV$ and $m_a>10\,\, \rm MeV$ differ only in their approach to thermalization, but the final state is determined by the same conditions: conservation of energy, conservation of radial momentum, and decoupling of bremsstrahlung. The first two conditions are enforced by simply determining the total energy and radial momentum density of the initial axion shell, and imposing it equal to the energy and radial momentum density of the relativistic fluid of photons and electrons. The three conditions determine the three quantities of the plasma immediately after thermalization, namely the Lorentz factor $\gamma$, the temperature in rest frame $T$, and the common chemical potential of photons and electrons $\mu/T$.

The determination of the final state of the plasma is simplified by making two assumptions, both of which are verified by the final result. First, since the thermalization is interrupted because the electron density is suppressed, we can already guess that the plasma at the end of thermalization will be dominantly composed of photons. Therefore, in its energy density and pressure, we may neglect the electron-positron contribution and use again the equation of state of relativistic Boltzmann particles with $p=\rho/3$. This means that Eqs.~\eqref{eq:energy_conservation},~\eqref{eq:momentum_conservation} and~\eqref{eq:bulk_velocity} are still satisfied. 

The second assumption that we make is that in the final state $\mu\gg T$; this implies that thermalization is interrupted much before chemical equilibrium is reached, so that $\mu$ does not lower significantly. With this assumption, photons behave as relativistic Boltzmann particles and electrons/positrons behave as non-relativistic Boltzmann particles. Taking the ratio of Eqs.~\eqref{eq:bremsstrahlung_equilibrium} and~\eqref{eq:momentum_conservation}, after replacing the cross section Eq.~\eqref{eq:nr_brem_cross} and the corresponding energy density and pressure for photons and number density for electrons/positrons, we find
\begin{equation}\label{eq:temperature_determination_numerical}
    \frac{e^{-m_e/T}}{T^3}=\frac{3\pi\gamma vr^2}{\sqrt{2} \alpha^3 \mathcal{P}}.
\end{equation}
One may obtain an approximate solution to this equation in the limit $T\ll  m_e$, which in turn requires the condition
\begin{equation}
    \frac{3\pi\gamma v r^2 m_e^3}{\sqrt{2} \alpha^3\mathcal{P}}\ll 1.
\end{equation}
In this limit, the solution is approximately
\begin{equation}\label{eq:temperature_determination_analytical}
T\sim\frac{m_e}{-\log\left[\frac{3\pi\gamma v r^2 m_e^3}{\sqrt{2}\alpha^3\mathcal{P}}\right]}.
\end{equation}
The conditions for applicability are not always verified in our region of interest, which is why for the results shown in the figures we numerically solve Eq.~\eqref{eq:temperature_determination_numerical}; however, Eq.~\eqref{eq:temperature_determination_analytical} has some pedagogical value, since it shows that the final temperature is a small fraction of the electron mass of order $10\%$ with weak dependence on the initial parameters of the fireball. One may compare this with the initial temperature of the plasma before bremsstrahlung sets it, in Eq.~\eqref{eq:initial_temperature_plasma}, which on the other hand is of the order of the initial average energy per particle in the comoving frame.

Finally, the chemical potential can be obtained from any of the equations Eqs.~\eqref{eq:momentum_conservation},~\eqref{eq:energy_conservation} and~\eqref{eq:bremsstrahlung_equilibrium}; for example, using Eq.~\eqref{eq:momentum_conservation} we find
\begin{equation}
    \frac{\mu}{T}=\log\left[\frac{32\gamma^2 v r^2 \Delta T^4}{\pi \mathcal{P}}\right].
\end{equation}
Numerically, we find that this quantity is indeed much larger than $1$ in all the region of fireball formation.

Fig.~\ref{fig:final_state} shows the properties of the plasma formed after thermalization. The plasma moves with a bulk Lorentz factor which is larger at lighter masses, due to the relativistic beaming in the photons emitted. The temperature reached is a fraction of the electron mass, with little dependence on the amount of energy injected; the chemical potential is larger close to the boundaries of the region, where the final state is farthest from the chemical equilibrium state. Consistently with our approximation scheme, the chemical potential never reaches values below about $8T$. Therefore, at the end of the thermalization phase, the plasma is composed of a dominant population of photons and a subdominant population of $e^\pm$, with a sub-MeV temperature.

\section{Fireball expansion}\label{sec:expansion}

The subsequent evolution is driven by radial expansion. As long as pair annihilation remains fast, this expansion happens in the hydrodynamical regime. In the case of $\gamma\gg1$, this regime is well-known from the analogous fireball formed in $\gamma$-ray bursts~\cite{Piran:1999kx}. In our parameter space, we find $\gamma$ to be always larger than $2.5$, which means that the assumption of ultra-relativistic motion is reasonably satisfied; this approach is all the more justified since the radial expansion increases the bulk kinetic energy at the expense of the internal energy, leading to a rapid increase of $\gamma$. Furthermore, since the number of particles cannot change in this phase of expansion, the average energy per particle (in the laboratory frame) is not significantly affected by the details of the expansion.

For $\gamma\gg1$, the thickness of the shell remains constant, since the fluid moves radially with the speed of light. The evolution of the thermodynamic quantities is ruled by the conservation laws of the entropy per particle, $\sigma_\gamma(r)/n_\gamma(r)$, where $\sigma_\gamma$ is the entropy density; the average energy per particle, $\gamma(r) \rho_\gamma(r)/n_\gamma(r)$, with $\rho_\gamma$ being the rest-frame energy density; and the total number of photons, $\gamma n_\gamma 4\pi r^2 \Delta$. Unprimed quantities refer to the frame comoving with the fluid. These conservation laws happen to be identical to the fireball equations when chemical equilibrium is maintained, as derived from the hydrodynamical equations, see, e.g., Ref.~\cite{Piran:1999kx}. We are assuming the plasma to be dominated by photons, so that the $e^\pm$ component can be neglected. The conservation of entropy per particle implies that the ratio $\mu/T$ remains constant, since for photons $\sigma_\gamma/n_\gamma$ is only a function of $\mu/T$. In turn, the conservation of the average energy per particle implies the constancy of the product $\gamma T$. Finally, the conservation of the photon number implies the scaling laws $\gamma\propto r$, and therefore $T\propto r^{-1}$. 

The expansion ends when pair annihilation runs out of equilibrium. At this point, photons moving at an angle $\theta$ with the radial direction will have a thermal distribution with an effective temperature $T'=T_f/\gamma_f(1-v_f\cos\theta)$, where the subscript $f$ means that the quantity must be evaluated at the moment of decoupling. The spectrum detected at Earth is given by the superposition of the emissivity from each point of the decoupling sphere.\footnote{This point had been overlooked in the original version of the manuscript, as well as in the published version. We submitted an Erratum to Physical Review D to clarify this point, on which Appendix~\ref{app:boosted_blackbody} is based.} We describe this calculation in Appendix~\ref{app:boosted_blackbody}, and show that the observable spectrum, in the instantaneous decoupling approximation, is non-thermal with a spectral shape
\begin{equation}
\frac{dN}{dE}\propto E e^{-\frac{E}{2\tau}},
\end{equation}
where $\tau=\gamma_f T_f=\gamma T$ due to the constancy of $\gamma T$ during the hydrodynamical expansion phase. As it should be, this quantity is of the same order of the average energy per particle in the lab-frame. The spectrum is therefore anti-pinched (also known as a Gamma spectrum) with an average energy $4\tau$ and pinch parameter $\alpha=1$.

\subsection{Remaining fraction of high-energy photons}

\begin{figure*}
    \centering
    \includegraphics[width=0.3\textwidth]{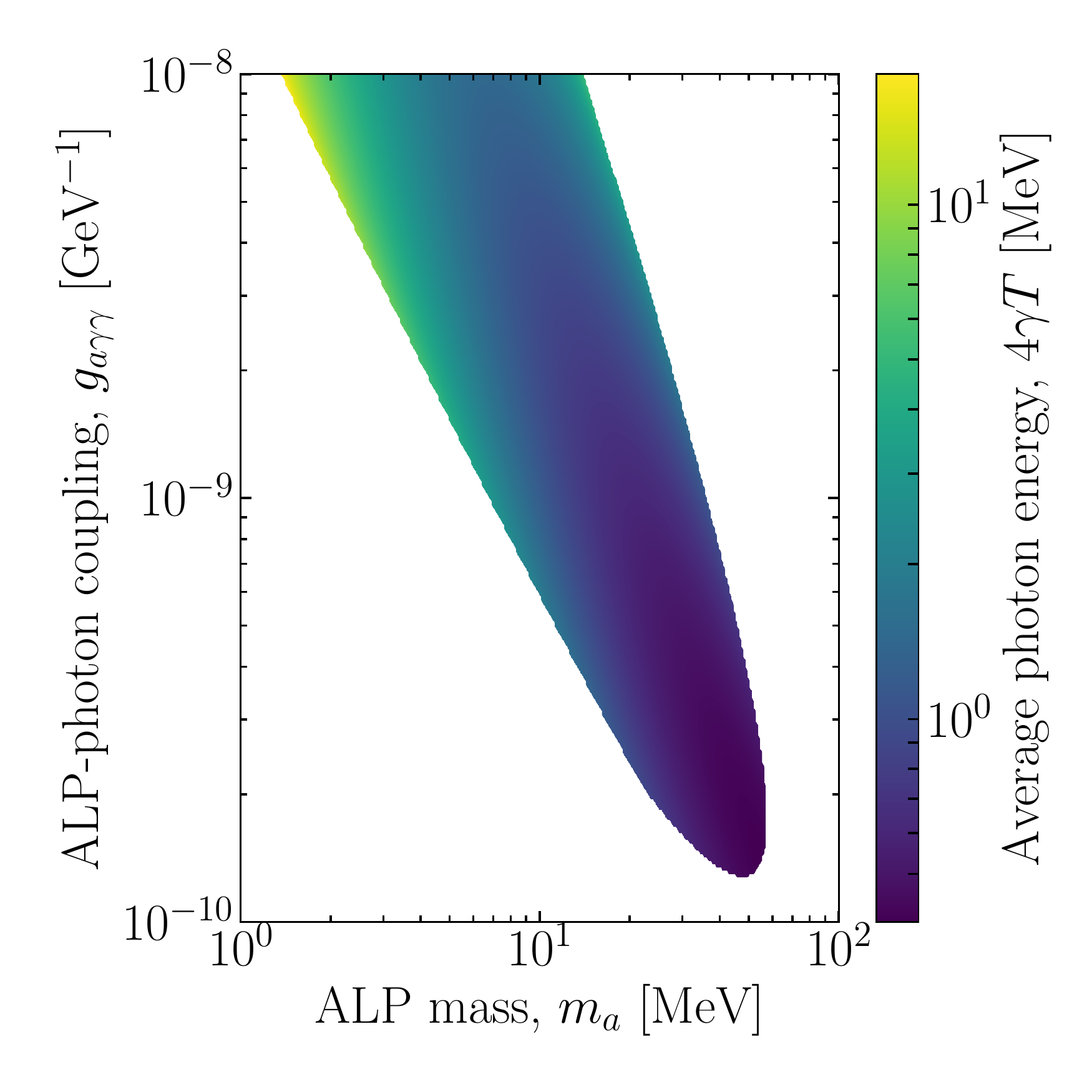}
    \includegraphics[width=0.3\textwidth]{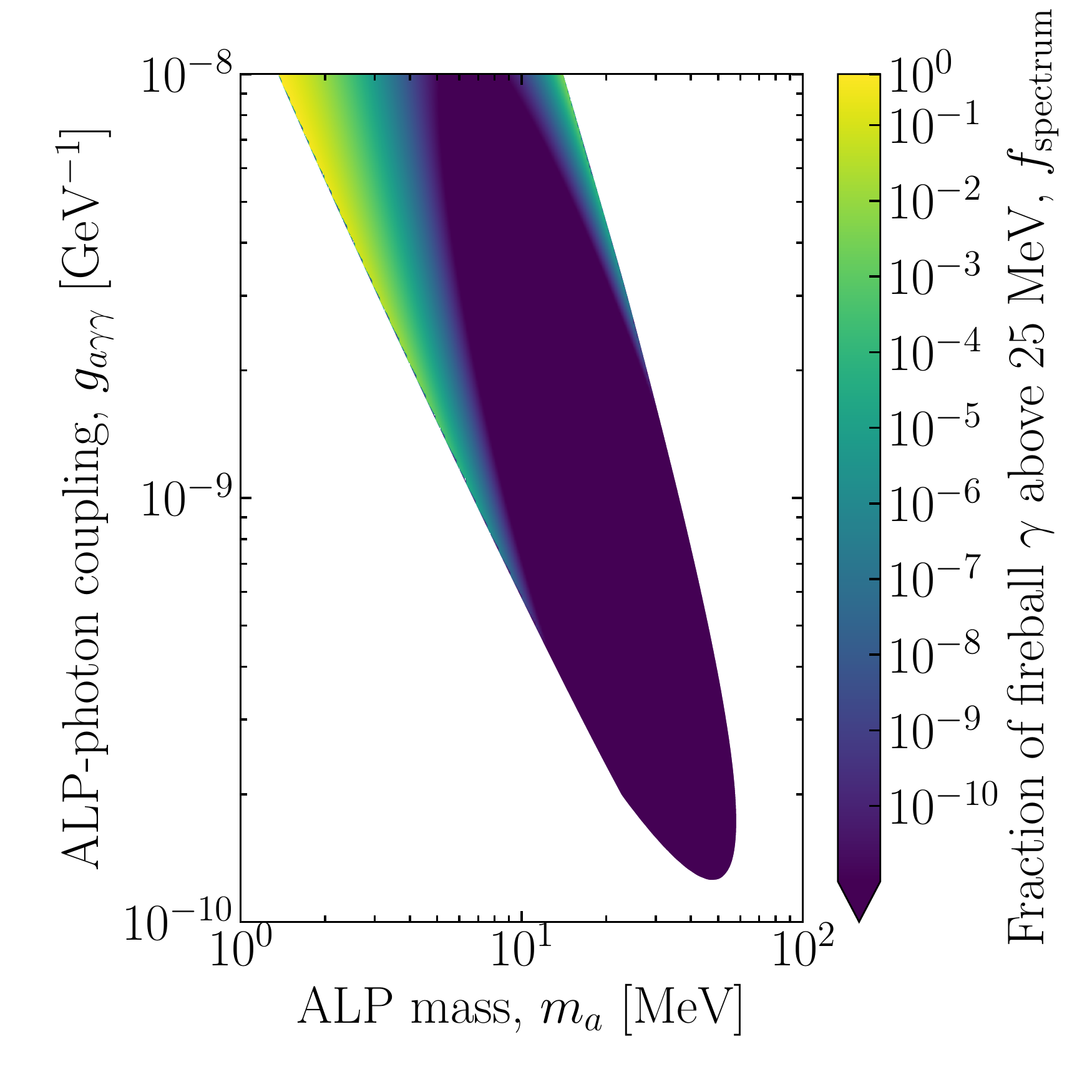}
    \includegraphics[width=0.3\textwidth]{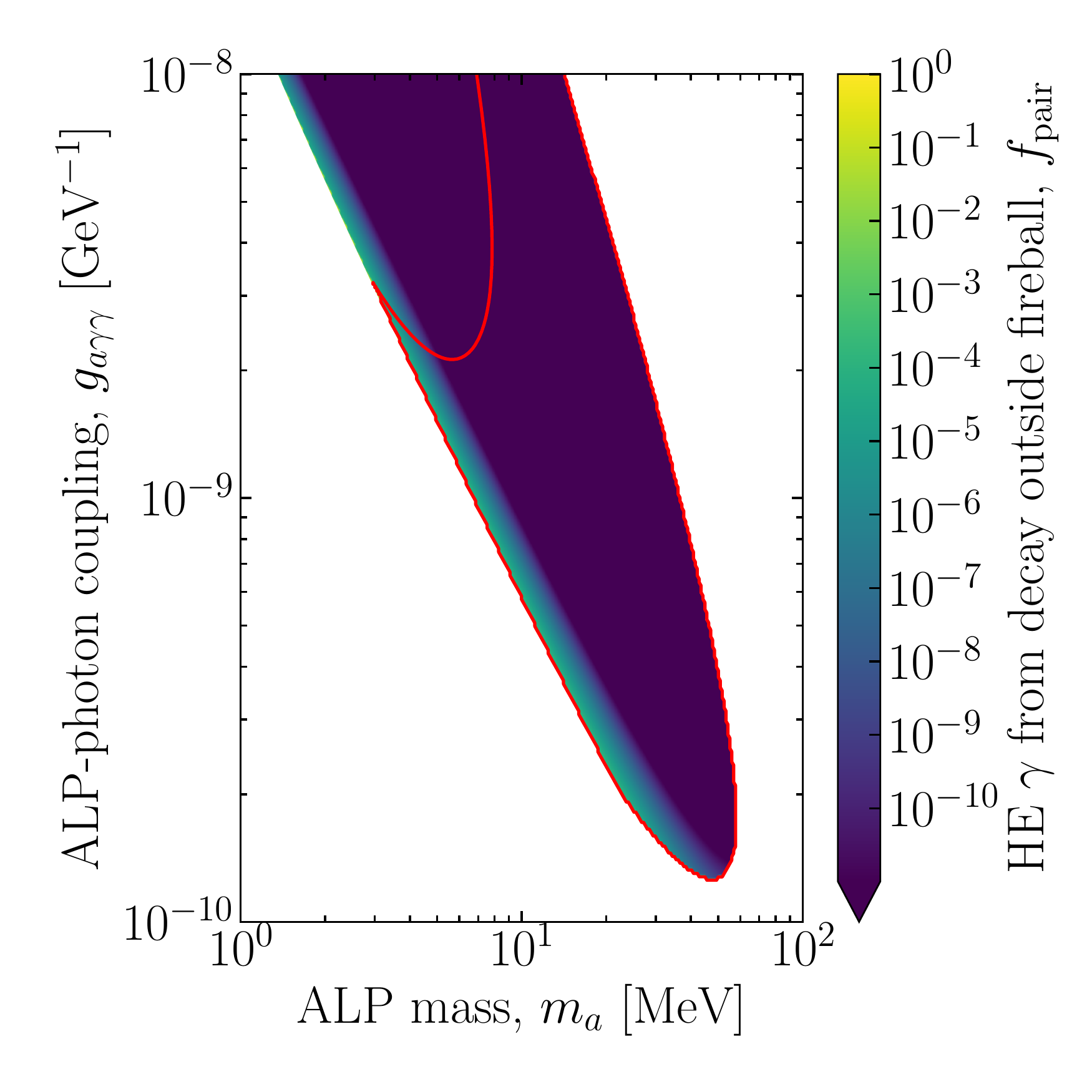}
    \caption{Spectral properties of the $\gamma$-ray emission from the fireball. We show the average energy of the thermalized photons (left), and the fraction of the thermalized photons integrated above $25\,\, \rm MeV$ (middle). We also show the fraction of axion decaying far enough from the center that the produced high-energy photons cannot thermalize and reach the Earth with their original energies; the red line is the boundary of the region in which the pair production reaction is always kinematically allowed from the criterion $E_\mathrm{CM,min}>m_e$.}
    \label{fig:average_energy}
\end{figure*}

Fig.~\ref{fig:average_energy} shows the average energy of the spectral distribution seen at Earth, namely $4\gamma T$, in the parameter space of interest. At low axion mass, the average energy can reach up to the order of MeV, mostly because of the higher bulk Lorentz factor due to the relativistic motion of the light axions. However, already at about $10\,\, \rm MeV$, the average energy becomes sub-MeV. To qualitatively understand the impact on the bounds from the non-observation of $\gamma$-rays in Refs.~\cite{Jaeckel:2017tud,Hoof:2022xbe}, we look at the fraction of photons that, after thermalization, remain at an energy higher than $25\, \rm MeV$, which is the threshold energy above which upper bounds on the fluence have been set. Therefore, we show in the middle panel of Fig.~\ref{fig:average_energy} the fraction of energy remaining in photons above $E_\mathrm{min}=25\,\, \rm MeV$, the minimum energy at which data were available from SN~1987A. Since the photon spectrum has the Gamma shape described above, namely proportional to $E e^{-E/2\gamma T}$, this fraction is
\begin{align}
    &f_\mathrm{spectrum}=\int_{E_\mathrm{min}}^{+\infty} e^{-E/2\gamma T}E^2 dE \bigg/\int_{0}^{+\infty} e^{-E/2\gamma T}E^2 dE \nonumber \\  &= e^{-E_\mathrm{min}/2\gamma T} \left[1+\frac{E_\mathrm{min}}{2\gamma T}+\frac{1}{2}\left(\frac{E_\mathrm{min}}{2\gamma T}\right)^2\right].
\end{align}
This fraction rapidly drops by more than 4 orders of magnitude at a mass larger than about $5\,\, \rm MeV$.

All the previous discussion relates to the bulk of the photons produced in the decay of the axion, at distances comparable with the mean decay length or the progenitor radius. However, at sufficiently large radii, the photons coming from the decay of the small fraction of surviving axions may not be able to thermalize with the dominant bath of lower energy photons, if the density of the photon shell has sufficiently rarefied. This small fraction of the total photon flux reaches Earth with the originally expected large energies.

To quantify the impact of this contribution, we estimate up to what radius the high-energy photons from late axion decays can thermalize with the bulk photon fluid. The dominant channel for thermalization is pair annihilation. While the temperature of the photon fluid is significantly lower than $m_e$, the photons from axion decay have a typical lab-frame energy $\bar{E}_a/2$, with $\bar{E}_a\sim 100\,\, \rm MeV$. In the rest frame of the decaying axion, the high-energy photons are isotropically distributed. Boosting to the laboratory frame, and then into the frame comoving with the plasma, we find that the center-of-mass energy for the collision of the two photons is
\begin{equation}
    E_\mathrm{CM}^2=3\gamma T \bar{E}_a\left(1-Vv\right), 
\end{equation}
where we have assumed an average energy $3T$ for the rest-frame photons in the plasma; here $V=\sqrt{1-\gamma^{-2}}$ is the bulk plasma velocity and $v=\sqrt{1-m_a^2/\bar{E}_a^2}$ is the typical axion velocity. As long as $E_\mathrm{CM}>m_e$, photons from axion decay are kinematically allowed to collide with the photons from the bulk of the plasma. Notice that the minimum value that $E_\mathrm{CM}^2$ can take as the plasma expands and $\gamma$ becomes larger is simply obtained by writing $V=1$
\begin{equation}
    E_\mathrm{CM,min}^2=3\gamma T\bar{E}_a\left(1-v\right).
\end{equation}
A sufficient condition for the reaction to be kinematically allowed is $E_\mathrm{CM,min}>m_e$.

The reaction will remain in equilibrium provided that
\begin{equation}
    \gamma n_\gamma \pi \alpha^2 \frac{\Delta}{E_\mathrm{CM}^2}\gg1,
\end{equation}
where we estimate the pair production cross-section as $\sigma_{\gamma\gamma\to e^+ e^-}\simeq \pi\alpha^2/s$, with $s$ the Mandelstam parameter of the collision. This estimate holds up to factors of order unity. Once the radius of the fireball has expanded by a factor $x$ to a radius $r_\mathrm{pair}=x r$, the condition for freeze-out of pair production reaction becomes
\begin{equation}
    \frac{n_{\gamma,0}\pi \alpha^2 \Delta}{3 T_0 \bar{E}_a x^2 \left[1-\sqrt{1-\frac{m_a^2}{\bar{E}_a^2}}\sqrt{1-\frac{1}{\gamma^2 x^2}}\right]}=1,
\end{equation}
where we identify by the suffix $0$ the quantities at the beginning of the expansion. Expanding $\sqrt{1-\frac{1}{\gamma^2 x^2}}\simeq 1-\frac{1}{2\gamma^2 x^2}$, we can obtain the factor of expansion as
\begin{equation}
    x^2=\frac{\frac{n_{\gamma,0}\pi\Delta \alpha^2}{3T_0 \bar{E}_a}-\frac{v}{2\gamma^2}}{1-v}.
\end{equation}

This defines the maximum radius $r_\mathrm{pair}=x r$ out of which injected photons are not able to thermalize with the bulk of the plasma. At this radius, the fraction of energy injected is exponentially suppressed as
\begin{equation}
    f_\mathrm{pair}=\frac{\int dE_a E_a e^{-r_\mathrm{pair}/\ell(E_a)} \frac{dN_a}{dE_a}}{\int dE_a E_a e^{-R/\ell(E_a)} \frac{dN_a}{dE_a}}.
\end{equation}
We show this suppression factor in the right panel of Fig.~\ref{fig:average_energy}: this should be interpreted as the fraction of energy produced by axions at a large enough radius that the decay photons are not able to thermalize with the remaining bath of low-energy photons. The red boundary identifies the region within which $E_\mathrm{CM,min}>m_e$, and therefore our treatment above is applicable. We define the fraction of energy which is still injected above the photon energy of $4\, \rm MeV$ as $f=\mathrm{max}\left[f_\mathrm{pair},\;f_\mathrm{spectrum}\right]$.

\section{Updated bounds from SN~1987A}\label{sec:bounds}

Since the current bounds from SN~1987A rely on the energy window above $25\,\, \rm MeV$~\cite{Jaeckel:2017tud,Hoof:2022xbe}, fireball formation can impact them. Unless the axion is very light and the photon energy remains large due to the large Lorentz factor, fireball formation reduces the photon flux above $25\,\, \rm MeV$ by more than ten orders of magnitude. (We comment on SMM data in the interval $4-25\,\rm MeV$ in Appendix~\ref{sec:above4}.)
This already indicates that a region of the parameter space, previously excluded by non-observation by SMM of the decay photons, may be ruled in---though, as we will see, this is not the case.

Thus, our first result is the region identified by the black solid line in Fig.~\ref{fig:final_bounds}, in which the bounds from both $\gamma$-ray decay from SN~1987A and diffuse supernova background need to be reevaluated, since the spectrum adopted in previous literature for $\gamma$-rays should actually be replaced by a Boltzmann spectrum with a temperature $2\gamma T$. Notice that this region is highly complementary to bounds from energy deposition in low-energy supernovae. If axions produced in the PNS travel and decay in the mantle, the ejecta kinetic energy becomes too large. Assuming that less than $0.1\,\rm B$ ($1\,\rm B \,(\rm bethe)=10^{51}\,\rm erg$) energy is deposited, as observed in low-energy supernovae, the solid red bounds are obtained~\cite{Caputo:2022mah}. A particularly conservative limit of $1\,\rm B$ leads to the dotted red bound. Since these bounds are calorimetric, they are not strongly affected---if at all---by the formation of a fireball. We notice that even within the red region, axion production has negligible impact on the inner dynamics of the SN core, where the axions themselves are produced. The cooling rate due to axion emission in this region of the parameter space is much lower than the one due to neutrino emission. Therefore, the axion flux we are considering is self-consistent.

 While the limits from SMM do not apply straightforwardly anymore, we find that the entire region of fireball formation is directly excluded by the non-observation of $\gamma$-rays at the Pierre Venus Orbiter (PVO). Launched in 1978 (last contact in 1992), PVO was part of NASA'S Pioneer Venus project and featured a $\gamma$-ray burst detector (OGBD) with two NaI photomultiplier detector units~\cite{colin197711,colin1980pioneer}. The sensitivity window of this detector was in a much lower energy range, between $0.2\, \rm MeV$ and $2\, \rm MeV$, which is just the region in which the average energy of the thermalized spectrum lies in the region of fireball formation. Moreover, PVO had a $4\pi$ acceptance and, being in orbit around Venus, it was obviously outside of the Earth's radiation belts, so the background was particularly low~\cite{Jaffe:1995sw}.
 
 As reported by Ref.~\cite{Jaffe:1995sw}, at the time of SN~1987A no excess over the background was observed in any of the energy bins observed by the instrument. While the response functions of the experiment are not published in this paper, we may extract a coarse bound on the fluence from Ref.~\cite{evans1979gamma}, which reports the detection of $\gamma$-ray bursts (GRB) in the same energy window with a fluence of the order of about $10^{-4}\,\rm erg\, cm^{-2}$. At the distance of SN~1987A of about $50\,\rm kpc$, this means that a minimum total energy $\mathcal{E}$ of about $3\times10^{43}\,\rm erg$ should have been detected. In the whole region of fireball formation, we find that the total energy $\mathcal{E}$ is always larger than $5\times 10^{48}\rm \, erg$, namely more than five orders of magnitude higher than the approximate exclusion. Therefore, even without refining the calculation, we can already claim that the entire region of fireball formation is excluded on the basis of the PVO non-observation of a $\gamma$-ray burst in the \textit{low} energy range $0.2-2\, \rm MeV$. In fact, the bounds from PVO may extend even outside the region of fireball formation, since the $\gamma$-ray spectrum from axion decay is more or less flat so that a non-negligible fraction of energy would be released in the low-energy window even without fireball formation; however, since in these regions bounds would be only complementary to the ones from SMM, we do not investigate this question further.
 
It is logical to ask if other detectors could probe the low-energy flux of the fireball from SN~1987A.\footnote{X-ray and $\gamma$-ray missions organized by launch date can be found in Ref.~\cite{LAUNCHES}.} While the GRS onboard of the SMM could detect photons even down to tens of keV~\cite{jones1985nineteenth}, no available public data exists of SN~1987A in this energy band. Other hard X-ray detectors which observed SN~1987A, such as a JPL balloon~\cite{1988ApJ...334L..81M} and the Ginga satellite~\cite{itoh1987x,Koshiba:1992yb}, did so only days after the collapse.

\section{Updated bounds from diffuse gamma-ray background}\label{sec:diffuse}
Photons originating from the decay of heavy axions produced by all past SNe would contribute to the diffuse cosmic $\gamma$-ray background. The idea of constraining neutrino radiative decays in this way dates back to the 1970s~\cite{Cowsik:1977vz}, and the argument has been revisited recently with applications to axions (see e.g.~\cite{Calore:2020tjw,Caputo:2021rux}).

We here follow closely Ref.~\cite{Caputo:2021rux} to compute the expected photon flux ignoring the possibility of a fireball. Assuming that all SNe occur at $z=1$, the photon energy flux (energy per unit area, time, and solid angle) due to axion decays is
\begin{equation}\label{eq:DiffusePrediction}
   \omega^2\frac{d\Phi_\gamma}{d\omega}=\frac{1}{4\pi}\zeta_a\,\frac{E_{\rm SN}}{E_{\rm av}}n_{\rm cc}\,
    \frac{2(T_{\rm eff}+2\omega) \omega^2}{T_{\rm eff}^2}\,e^{-2\omega/T_{\rm eff}},
\end{equation}
where $\zeta_a$ is a fudge factor proportional to $g_{a\gamma\gamma}^2$, $E_{\rm SN}=3\times 10^{53}\,\rm erg$ is the typical energy released by a SN, $T_{\rm eff}$ is the effective temperature akin to the one appearing in Eq.~\eqref{eq:hoof}, $E_{\rm av}=3T_{\rm eff}$ is the axion average energy, and $n_{\rm cc}\simeq 10^{7} \,\rm Mpc^{-3}$ is the core-collapse number
per comoving volume per redshift interval. This spectrum has a maximum at $\omega_{\rm max}=T_{\rm eff}(1+\sqrt3)/2\simeq 1.37\,T_{\rm eff}$.

The diffuse $\gamma$ flux produced by axions should be compared with the measurements of the extragalactic background radiation~\cite{Fermi-LAT:2014ryh}. For axions with masses and couplings not allowing the formation of a fireball, we can use the range of the extragalactic background spectrum 2--200~MeV~\cite{Fermi-LAT:2014ryh},
\begin{equation}\label{eq:DiffuseFlux}
    \omega^2\frac{d\Phi_\gamma^{\rm observed}}{d\omega}\Big|_{2-200\,\rm MeV}\simeq 2\times10^{-3}~{\rm MeV}~{\rm cm}^{-2}~{\rm s}^{-1}~{\rm ster}^{-1}.
\end{equation}
Therefore, the bound is found comparing Eq.~\eqref{eq:DiffusePrediction} evaluated at $\omega_{\rm max}$,
\begin{align}
\omega^2\frac{d\Phi_\gamma}{d\omega}\Big|_{\rm max}&=\zeta_a\,E_{\rm SN}n_{\rm cc}\,\frac{7+4\sqrt3}{12\pi}\,e^{-(1+\sqrt3)}
  \nonumber\\[1ex]
  &=\zeta_a\,\frac{n_{\rm cc}}{10^7\,\rm Mpc^{-3}}\,46.2~{\rm MeV}~{\rm cm}^{-2}~{\rm s}^{-1}~{\rm ster}^{-1},
\end{align}
with Eq.~\eqref{eq:DiffuseFlux}. One obtains~\cite{Caputo:2021rux}
\begin{equation}\label{eq:genericgammalimit}
    \zeta_a\alt 0.43\times10^{-4}\big/n_{\rm cc},
\end{equation}
independently of the assumed average energy of the emitted bosons. This corresponds to a constraint on axions with a photon coupling $g_{a\gamma\gamma}\lesssim 10^{-10} \, \rm GeV^{-1}$~\cite{Caputo:2021rux,Caputo:2022mah}. The result does not depend strongly on the assumed redshift dependence of the cosmic core-collapse rate $n_{\rm cc}$.

The formation of the fireball impacts the diffuse $\gamma$-ray bounds: as the average energy of photons is driven below $1\,\, \rm MeV$, the flux from axion decays needs to be compared to a much larger astrophysical background (see e.g. Figure 10 of Ref.~\cite{Fermi-LAT:2014ryh}). A very conservative estimate can be obtained using the largest value of the extragalactic background light energy flux,
\begin{equation}\label{eq:DiffuseFlux2}
    \omega^2\frac{d\Phi_\gamma^{\rm observed}}{d\omega}\Big|_{\lesssim 100\,\rm keV}\lesssim 5\times10^{-2}~{\rm MeV}~{\rm cm}^{-2}~{\rm s}^{-1}~{\rm ster}^{-1}.
\end{equation}
This is 25 times larger than the value that is observed in the 2–200 MeV energy range. Therefore, the diffuse $\gamma$-ray bound is relaxed at most by a factor of 5 ($g_{a\gamma\gamma}\lesssim 5 \times 10^{-10}\,\rm GeV^{-1}$) in the region corresponding to fireball formation. On the other hand, the bound rapidly  degrades at $m_a\simeq 50\,\, \rm MeV$~\cite{Caputo:2022mah}, so we cannot commit to a precise evaluation of this region, and a dedicated analysis is needed to revisit this bound at large masses.

\section{Discussion and outlook}\label{sec:discussion}

Heavy axions coupling to photons can be produced in the hot core of supernovae and decay back to photons with around $100\,\, \rm MeV$. While it has always been assumed that such photons would travel freely, in part of the axion parameter space photons form a fireball, and their energy degrades to values below $1\,\, \rm MeV$. Part of the parameter space previously thought to be excluded by observations of SN~1987A with SMM~\cite{Jaeckel:2017tud,Hoof:2022xbe} is actually excluded by PVO data~\cite{PVO,Jaffe:1995sw}, as the energy of the photons at Earth would have been much lower than expected. 
PVO observations, previously applied only to neutrino radiative decays~\cite{Jaffe:1995sw}, are relevant since cosmological constraints apply in the region of interest only for a large reheating temperature~\cite{Depta:2020zbh,Langhoff:2022bij}.
Constraints arising from the extragalactic background light (see e.g. Ref.~\cite{Caputo:2021rux} and references therein) also get relaxed. Other complementary probes in the regions might be, for example, very low luminosity SNe and their light-curve shape and spectral line
velocities, that can potentially probe unexplored parts of the parameter space~\cite{Caputo:2022mah}. Another interesting probe could rely on past~\cite{LIGOScientific:2017ync} and future neutron star merger observations, which we plan to explore in forthcoming work. 
We stress that our discussion applies to different models, such as axions coupling to charged leptons~\cite{Ferreira:2022xlw,Caputo:2021rux} and heavy neutral leptons featuring a dipole portal~\cite{Brdar:2023tm}. Therefore, constraints arising from SN $\gamma$-ray observations should be updated to include PVO data. Fireball formation is even easier for axions with an electron coupling, since they can decay to electron-positron pairs and bremsstrahlung can drive the thermalization without the need of producing pairs through two-photon annihilation. 

\bigskip

\acknowledgments

We thank Hans-Thomas Janka, Georg Raffelt, and Irene Tamborra for comments on the first draft of this paper.
 MD acknowledges the support of the National Sciences and Engineering Research Council of Canada (NSERC). 
 DF is supported by the {\sc Villum Fonden} under project no.~29388.   This project has received funding from the European Union's Horizon 2020 research and innovation program under the Marie Sklodowska-Curie grant agreement No.~847523 `INTERACTIONS'.  
 GMT acknowledges support by the National Science Foundation under Grant Number PHY-2210361 and by the US-Israeli BSF Grant 2018236. EV acknowledges support by the European
Research Council (ERC) under the European Union's Horizon Europe research and innovation programme (grant agreement No. 101040019). Views and opinions expressed are however those of the author(s) only and do not necessarily reflect those of the European Union.  This work used resources provided by the High Performance Computing Center at the University of Copenhagen. 

\appendix


\section{Observable photon spectrum from the fireball}\label{app:boosted_blackbody}

\begin{figure}
    \centering
\includegraphics[width=1.\columnwidth]{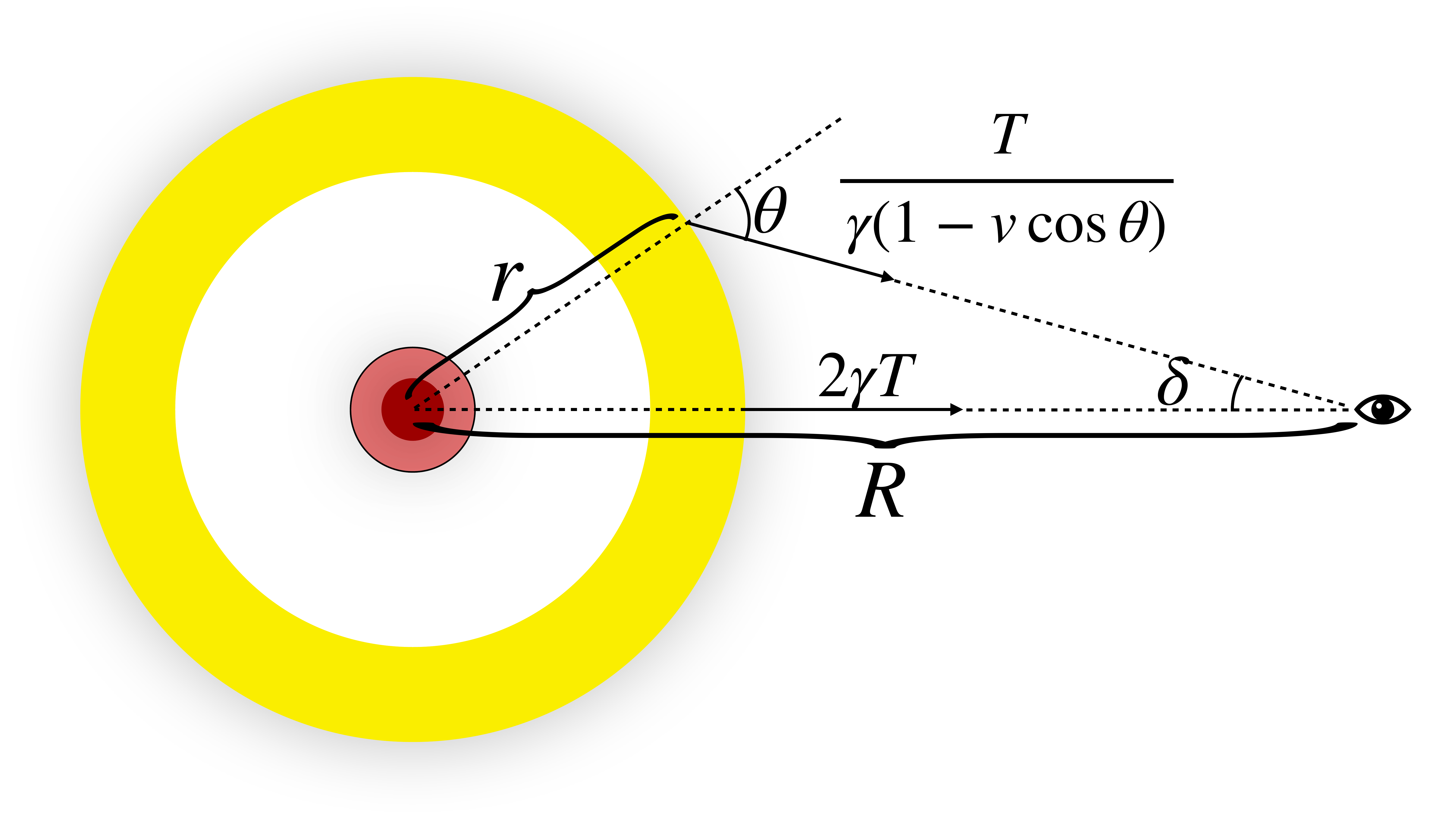}
    \caption{Schematic view of the effective temperature at each point of the decoupling sphere.}
    \label{fig:erratum}
\end{figure}

In this appendix we discuss the spectrum observable at Earth from the fireball after the photons have decoupled. At the moment of decoupling, the distribution function of the photons at each point of the decoupling sphere is 
\begin{equation}
f(E,\theta)=\frac{1}{\exp\left[\frac{\gamma E}{T}\left(1-v\cos\theta\right)+\eta\right]-1},
\end{equation}
where $\theta$ is the angle with the radial direction. We denote by $r$ the radius of the decoupling sphere, and $R$ the radius of observation, and we assume $r\ll R$. The viewing angle under which the observer sees a point of the decoupling sphere $\delta \ll 1$ is connected by the escaping angle $\theta$ approximately by the geometric relation
\begin{equation}
\delta\simeq\frac{r}{R}\sin\theta.
\end{equation}
The observer sees from every direction a blackbody distribution function with a direction-dependent temperature. If the source is distant enough, then the different directions cannot be resolved, and only the angle-integrated distribution function has meaning. This is
\begin{equation}
\frac{dN}{dt dS E^2 dE}=\int \frac{\sin \delta \cos\delta d \delta 2\pi}{(2\pi)^3} f\left[E,\theta(\delta)\right]
\end{equation}
for each degree of freedom (for a photon we should thus multiply by $2$ to account for the two polarizations).
After calling $\cos\theta=y$, which runs from $0$ to $1$ (since only outgoing photons can escape the decoupling surface), this becomes
\begin{equation}
\frac{dN}{dt dSE^2 dE}=\frac{r^2}{4\pi^2R^2}\int_0^1\frac{y}{\exp\left[\frac{\gamma E}{T}\left(1-vy\right)+\eta\right]-1}dy.
\end{equation}
Notice that if we integrate this expression over $E^3 dE$ to obtain the energy flux, we obtain
correctly the energy flux emitted \textit{by the forward modes only} in a boosted blackbody spectrum from a sphere of radius $r$, scaled by the factor $(r/R)^2$ at a distance $R$, so this expression conserves energy as it should.

This integral can be done exactly in the limit that $v\simeq 1$, which is usually the case, since the fireball expands and develops large radial bulk velocity before decoupling. However, during the fireball expansion, also the temperature goes down as $T\propto \gamma^{-1}$. Therefore, we first express everything in terms of the effective temperature $\tau=\gamma T$ and take the limit as $\gamma\to \infty$
\begin{equation}
\frac{dN}{dt dSE^2 dE}=\frac{r^2}{4\pi^2R^2}\int_0^1\frac{y}{\exp\left[\frac{E y}{2\tau}+\frac{E \gamma^2}{\tau}(1-y)+\eta\right]-1}dy.
\end{equation}

This integral can be performed explicitly in terms of polylogarithms. However, when keeping only the leading order terms in $\gamma^{-2}$, the result is very simple
\begin{equation}
\frac{dN}{dt dSE^2 dE}=-\frac{r^2}{4\pi^2R^2} \frac{\tau}{E \gamma^2}\log\left[1-e^{-\eta-\frac{E}{2\tau}}\right].
\end{equation}
Somewhat satisfyingly, the result does not now depend on the decoupling radius, since the combinations $\gamma/r$ and $\tau$ are both constant during the free expansion phase. Generalizing to the case of fermions, which is not directly interesting here, we find a spectrum
\begin{equation}
\frac{dN}{dt dSE^2 dE}=\frac{r^2}{4\pi^2R^2} \frac{\tau}{E \gamma^2}\log\left[1+e^{-\eta-\frac{E}{2\tau}}\right].
\end{equation}
If we integrate the spectrum to obtain the energy flux, we find that in both cases this is exactly equal to $4\gamma^2 \rho /3$, namely the energy flux of a relativistic fluid in ultra-relativistic motion (in this case we do not have to correct for only the forward modes being emitted, since for $v\simeq 1$ all energy is contained in the forward modes).

For the purposes of the bounds, we are mainly interested in the spectral shape, which at this point is not a blackbody spectrum, but rather a spectrum of the form
\begin{equation}
\frac{dN}{dE}\propto -E \log\left[1-e^{-\eta-\frac{E}{2\tau}}\right].
\end{equation}
In the limit of large degeneracy, which is the relevant one for the fireball formed outside of a supernova, this becomes
\begin{equation}
\frac{dN}{dE}\propto E e^{-\frac{E}{2\tau}},
\end{equation}
which has the form of an anti-pinched spectrum (also known as a Gamma spectrum) with an average energy $4\tau$ and pinch parameter $\alpha=1$.

Notice that the conclusion that the average energy is $4\tau$ follows from simple arguments of relativistic invariance. Since the energy flux reaching the Earth $\left(dE/dtdS\right)_{\oplus}$ is simply the energy flux $\left(dE/dtdS\right)_\mathrm{SN}$ scaled by the squared radii
\begin{equation}
    \left(\frac{dE}{dtdS}\right)_{\oplus}=\left(\frac{r}{R}\right)^2 \left(\frac{dE}{dtdS}\right)_\mathrm{SN},
\end{equation}
and the same relation holds for the number fluxes, it follows that the average energy is simply equal to the ratio
\begin{equation}
    \langle E \rangle=\frac{\left(\frac{dE}{dtdS}\right)_\mathrm{SN}}{\left(\frac{dN}{dtdS}\right)_\mathrm{SN}}.
\end{equation}
Using now conventional Lorentz boosting we have
\begin{equation}
    \left(\frac{dE}{dtdS}\right)_\mathrm{SN}=\frac{4}{3}\gamma_f^2 v_f \rho_f,
\end{equation}
with $\rho_f$ the rest-frame energy density at decoupling, and
\begin{equation}
    \left(\frac{dN}{dtdS}\right)_\mathrm{SN}=\gamma_f n_f v_f,
\end{equation}
from the relation valid for the Maxwell-Boltzmann limit $\rho_f/n_f=3T_f$ we find
\begin{equation}
    \langle E \rangle =\frac{4}{3} 3\gamma_f T_f=4\tau.
\end{equation}

\begin{figure}
    \centering
\includegraphics[width=0.5\textwidth]{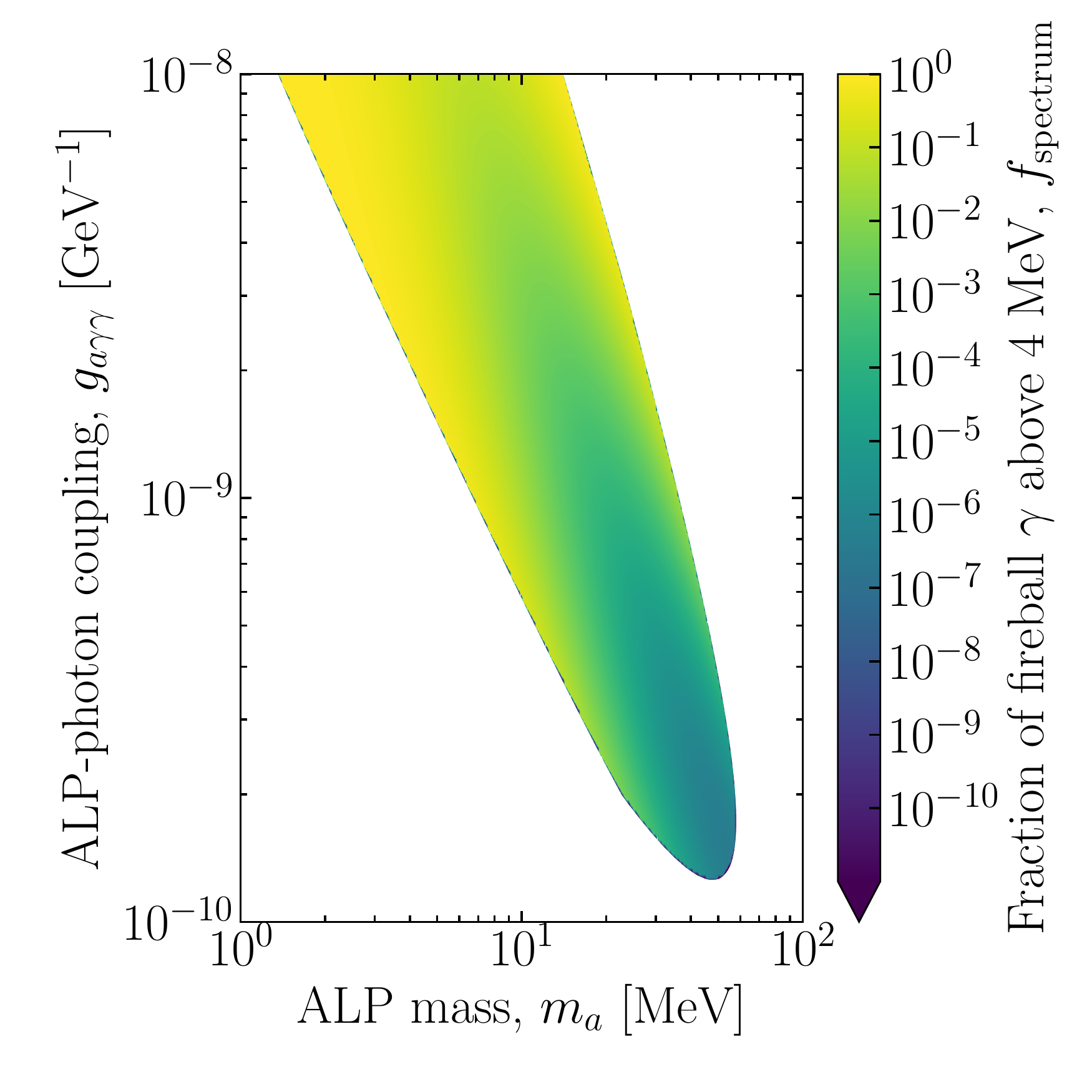}
    \caption{Fraction of the thermalized photons integrated above $4\,\, \rm MeV$.}
    \label{fig:fraction_4_meV}
\end{figure}

\section{Best reach of Solar Maximum Mission}\label{sec:above4}

In the main text we have discussed what is the fraction of photons remaining above $25$~MeV, since Refs.~\cite{Jaeckel:2017tud,Hoof:2022xbe} both use $\gamma$-ray data above this energy. In reality, SMM did have additional observations down to about $4$~MeV. Therefore, it makes sense to also look at the fraction of photons remaining above $4$~MeV; this would provide a guide as to how powerful the bounds from SMM could be by accounting also for these additional data which were not used in past approaches.

Fig.~\ref{fig:fraction_4_meV} shows this fraction of energy. We find this to be significantly higher than the fraction of energy injected above $25$~MeV, which is expected since the tail of the spectrum decreases exponentially. Therefore, SMM may in principle still constrain a significant part of the region of fireball formation, though with different data and a different approach than what has been done in the past. However, we do not follow up on this question further, since the bounds from PVO robustly exclude the entirety of the region.

\bibliographystyle{bibi}
\bibliography{References}

\onecolumngrid
\appendix

\clearpage

\setcounter{equation}{0}
\setcounter{figure}{0}
\setcounter{table}{0}
\setcounter{page}{1}
\makeatletter
\renewcommand{\theequation}{S\arabic{equation}}
\renewcommand{\thefigure}{S\arabic{figure}}
\renewcommand{\thepage}{S\arabic{page}}


\end{document}